\documentclass{article}
\usepackage{graphicx} 
\usepackage{amsmath}
\usepackage{amssymb}
\usepackage{caption}
\usepackage{float}
\usepackage{cite}
\usepackage{subfigure}
\usepackage{geometry}
\title{Tsallis holographic dark energy with power law ansatz approach}
\author{%
    Oem Trivedi $^{1}$\thanks{oem.t@ahduni.edu.in}, 
    Maxim Khlopov$^{2,3,4}$\thanks{khlopov@apc.in2p3.fr}, 
    Alexander V. Timoshkin$^{5,6}$ \thanks{alex.timosh@rambler.ru}
}

\date{%
    \small
    $^1$International Centre for Space and Cosmology, School of Arts and Sciences, Ahmedabad University, Ahmedabad-380009, Gujarat, India\\%
	$^{2}$Research Institute of Physics, Southern Federal University, 344090 Rostov-on-Don, Russia\\
    $^{3}$Virtual Institute of Astroparticle Physics, 75018 Paris, France\\
    $^{4}$Center for Cosmoparticle Physics Cosmion, National Research Nuclear University “MEPHI”, 115409 Moscow, Russia\\
    $^{5}$Institute of Scientific Research and Development,Tomsk State Pedagogical University (TSPU), 634061 Tomsk, Russia\\%
    $^{6}$ Lab. for Theor. Cosmology, International Centre of Gravity and Cosmos,
Tomsk State University of Control Systems and Radio Electronics (TUSUR), 634050 Tomsk, Russia\\%
	\today
}

\begin{document}

\maketitle

\begin{abstract}
    Holographic principles have proven to be a very interesting approach towards dealing with the issues of the late-time acceleration of the universe, which has resulted in a great amount of work on holographic dark energy models. We consider one such very interesting holographic scenario, namely the Tsallis Holographic dark energy model and consider an ansatz based approach to such models. We consider three cosmological scenarios in such models, namely those with viscous , non-viscous and Chaplygin gas scenarios, discussing various crucial aspects related to these models. We discuss various crucial properties of the Tsallis model in such scenarios and see how the phantom divide is crossed in each case, but it's only the Chaplygin gas models which give a better view on stability issues.  
\end{abstract}

\section{Introduction}
The surprising revelation of the late-time acceleration of the Universe startled the cosmology community \cite{SupernovaSearchTeam:1998fmf}. Subsequent efforts have been extensive in elucidating this expansion phenomenon. Numerous approaches have been pursued, encompassing traditional avenues like the Cosmological constant \cite{Weinberg:1988cp,Lombriser:2019jia,Padmanabhan:2002ji}, as well as unconventional theories such as Modified gravity \cite{Capozziello:2011et,Nojiri:2010wj,Nojiri:2017ncd} and models involving scalar fields driving late-time cosmic acceleration \cite{Zlatev:1998tr,Tsujikawa:2013fta,Faraoni:2000wk,Gasperini:2001pc,Capozziello:2003tk,Capozziello:2002rd,Odintsov:2023weg}. Additionally, diverse perspectives within quantum gravity have contributed to addressing the cosmic-acceleration enigma, ranging from Braneworld cosmology in string theory to theories like loop quantum cosmology and asymptotically safe cosmology \cite{Sahni:2002dx,Sami:2004xk,Tretyakov:2005en,Chen:2008ca,Fu:2008gh,Bonanno:2001hi,Bonanno:2001xi,Bentivegna:2003rr,Reuter:2005kb,Bonanno:2007wg,Weinberg:2009wa,Trivedi:2022svr,Trivedi:2022ipa}. Nevertheless, these endeavors have underscored certain discrepancies, notably the Hubble tension, which refers to disparities in the values of the Hubble constant derived from detailed CMB maps, combined with Baryon Acoustic Oscillations data and Supernovae Type Ia (SNeIa) data \cite{Planck:2018vyg,riess2019large,riess2021comprehensive}. Consequently, the current epoch of the universe presents a myriad of inquiries and appears poised to evolve into a domain where advanced gravitational physics will illuminate a deeper comprehension of cosmology.
\\
\\
Among the various proposed solutions addressing the dark energy problem , one noteworthy proposition is the Holographic principle \cite{tHooft:1993dmi,Susskind:1994vu}, which carries significance within the realm of quantum gravity. At its core, the holographic principle posits that a system's entropy isn't dictated by its volume but rather by its surface area \cite{Bousso:1999xy}. Cohen et al. \cite{Cohen:1998zx}, drawing from this principle, proposed a linkage between short-distance and long-distance cutoffs in quantum field theory, attributing it to the constraints imposed by black hole formation. In essence, if $\rho$ denotes the quantum zero-point energy density due to a short-distance cutoff, the energy within a region of size $L$ should not surpass the mass of a black hole of equivalent size, leading to the inequality $L^3\rho\leq LM_{pl}^2$. The maximum permissible value for the infrared cutoff ($L_{_{IR}}$) precisely satisfies this inequality, yielding the relationship:
\begin{equation}\label{simphde}
\rho=3M_{p}^2c^2L_{_{IR}}^{-2},
\end{equation}
where $c$ represents an arbitrary parameter, and $M_{p}$ signifies the reduced Planck mass.

The holographic principle has found extensive application in cosmology, particularly in elucidating the late-time dark energy era, commonly known as holographic dark energy (HDE) (for a comprehensive review, refer to \cite{Wang:2016och}). In this framework, the infrared cutoff, $L_{_{IR}}$, originates from cosmological considerations. Nojiri et al. \cite{Nojiri:2005pu, Nojiri:2017opc,Nojiri:2021iko,Nojiri:2020wmh} introduced the most general form of this cutoff, termed the generalized HDE, which encompasses various combinations of the FRW parameters including the Hubble constant, particle and future horizons, cosmological constant, and the finite lifetime of the universe. Numerous other studies have investigated holographic dark energy from diverse perspectives in recent years \cite{Nojiri:2017opc,Oliveros:2022biu,Granda:2008dk,Khurshudyan:2016gmb,Wang:2016och, Khurshudyan:2016uql,Belkacemi:2011zk, Zhang:2011zze,Setare:2010zy,Nozari:2009zk,Sheykhi:2009dz, Xu:2009xi,Wei:2009au,Setare:2008hm,Saridakis:2007wx,Setare:2006yj, Felegary:2016znh,Dheepika:2021fqv,Brevik:2022aqp,Brevik:2023vaf,Nojiri:2022aof,Nojiri:2019skr}. In the present work we want to discuss the properties of the Tsallis HDE model in a scale factor ansatz based approach, which has been quite heavily discussed in recent times \cite{Pasqua:2014oya,Chattopadhyay:2013mta,Chattopadhyay:2014yda,Jawad:2016tne}. In the next section we shall give a brief overview of the cosmological settings in which we would like to study our HDE scenario, discussing the non-viscous, viscous and Chaplygin gas EOS in detail and their implications on the evolution of the universe. In section III we discuss the important aspects of the models we are considering and see which of these can be best placed in the face of the usual issues faced by holographic models. We conclude our work in section IV. 
\\
\\
\section{Tsallis Holographic dark energy and the various EOS }
One particularly intriguing proposition within the realm of Holographic Dark Energy is the Tsallis Holographic Dark Energy model, often abbreviated as THDE. Tsallis and Cirto \cite{Tsallis:2012js} introduced a generalized form of entropy, famously known as Tsallis entropy, to address thermodynamic inconsistencies in non-standard systems, such as black holes. The pioneering investigations into dark energy models using Tsallis' non-extensive statistical framework can be traced back to \cite{Barboza:2014yfe}, with further explorations in cosmology detailed in \cite{Nunes:2015xsa}. This type of entropy aligns well with the Friedmann equations and Padmanabhan's proposal regarding the emergence of spacetime \cite{moradpour2016implications}.

Similar to the conventional HDE model, it is feasible to formulate dark energy models utilizing Tsallis entropy. As a result, Tsallis Holographic Dark Energy (THDE), using the Hubble horizon as an infrared (IR) cutoff, was introduced in \cite{Tavayef:2018xwx}. Building upon this foundation, the dynamics of Friedmann-Robertson-Walker (FRW) universes, considering dark matter and THDE with various IR cutoffs such as the apparent horizon, the particle horizon, the Ricci scalar curvature scale, and the Granda-Oliveros (GO) scale, were examined in non-interacting and interacting scenarios \cite{zadeh2018note,aly2019tsallis,srivastava2020tsallis,sharma2021tsallis}.

It was found that the THDE model with the particle horizon as the IR cutoff provides an explanation for the ongoing accelerated expansion of the universe, unlike the corresponding conventional HDE model. The findings presented in \cite{zadeh2018note} indicate that the stability of the THDE model varies depending on the choice of the GO scale and the Ricci scalar cutoff, in both interacting and non-interacting scenarios. However, in \cite{aly2019tsallis}, it is demonstrated that the THDE model with the GO scale as the IR cutoff remains stable in an $(n+1)$-dimensional FRW universe. For the Tsallis HDE scenario, the horizon entropy is given by \begin{equation} \label{stsa}
    S_{h} = \gamma r_{h}^{2 \sigma}
\end{equation}
where $\gamma_{t}$ is a constant in terms of the Planck area and the Tsallis parameter is $\sigma$. Using the Tsallis entropy, one can write the Holographic DE energy density as \begin{equation}
    \rho = \frac{3 c^2}{R_{h}^{4-2 \sigma}}
\end{equation}
where $R_{h}$ here refers the infrared cutoff scale, similar to the one in the simple holographic energy density. Furthermore, we shall be considering that there is no interaction happening between the dark energy and dark matter sectors in this work. In the non-viscous scenario, the equation of state is the usual $p = w \rho $, where $w$ is the equation of state parameter. The equation of state that we consider for the viscous fluid configuration is given by \cite{Nojiri:2006zh,Timoshkin:2023pfb} \begin{equation}
    p = w \rho - 3 \epsilon_{0} H
\end{equation}
where $\epsilon_{0} $ is a thermodynamic parameter which can be considered to be either time dependent or time independent. In this scenario, we consider dark energy to be a viscous holographic dark fluid while next, we will also be considering a generalized Chaplygin gas model \cite{Bento:2002ps}, characterized by the EOS \begin{equation}
    p = - \frac{A}{\rho^{\alpha}}
\end{equation}
where A and $\alpha$ are assumed to be positive constants, where for $\alpha = 1$, one obtains the usual Chaplygin gas model. In this scenario, we shall be considering dark energy to be a holographic Chaplygin gas. Furthermore, we shall be considering an ansatz for the scale factor of the power law form as follows \cite{Nojiri:2003jn,Nojiri:2003vn,Nojiri:2004ip,Nojiri:2004pf,Nojiri:2005pu,Nojiri:2005sr,Nojiri:2003ft} \begin{equation} \label{scalefac}
    a(t) = a_{0} (t_{s} - t )^n
\end{equation}
The above form of the scale factor has been used in various seminal works to understand late time acceleration of the universe and even cosmological singularities ( see \cite{Trivedi:2023zlf,deHaro:2023lbq} for recent reviews on the same) by Odintsov, Nojiri, Barrow etc and is fit to be used for our purposes too as it accounts for a lot of interesting cosmological epochs. Such an ansatz based approach to Holographic dark energy has been recently pursued in other works too \cite{pasqua2014study,Pasqua:2014oya,Chattopadhyay:2013mta,Chattopadhyay:2014yda}. 
\\
\\
\section{Analysis of the various scenarios}
\subsection{Non-viscous fluid case}
Choosing the conventional holographic dark energy density :
\begin{equation} \label{1}
\rho_{\Lambda}=\frac{3c^2}{R_{h}^2}
\end{equation}
where \(c\) is a constant, and choosing the scale to correspond with the event horizon, with the \(R_h\) being the future event horizon given by:
\begin{equation} \label{eventrh}
R_h= a\int_t^\infty \frac{dt}{a}=a\int_a^\infty\frac{da}{Ha^2}
\end{equation}
The critical energy density, \(\rho_{cr}\), is given by the following relation:
\begin{equation}
\rho_{cr}=3H^2
\end{equation}
Now we define the dimensionless dark energy as:
\begin{equation} \label{omega}
\Omega_{\Lambda}=\frac{\rho_{\Lambda}}{\rho_{cr}}=\frac{c^2}{R_{h}^2H^2}
\end{equation}
Using the definition of \(\Omega_\Lambda\) and the relation for \(\dot{R_{h}}\) from equations \eqref{eventrh}-\eqref{omega}, we have:
\begin{equation}
\dot{R_{h}} = R_{h}H-1=\frac{c}{\sqrt{\Omega_\Lambda}}-1
\end{equation}
By considering the definition of holographic energy density \(\rho_{\rm \Lambda}\) and using the expressions for \(\Omega_\Lambda\) and \(\dot{R_{h}}\), we can find:
\begin{equation}
\dot{\rho_{\Lambda}}=-2H\left(1-\frac{\sqrt{\Omega_\Lambda}}{c}\right)\rho_{\Lambda}
\end{equation}
Substituting this relation into the following equation:
\begin{equation}
\dot{\rho}_{\rm \Lambda}+3H(1+w_{\rm \Lambda})\rho_{\rm \Lambda} =0
\end{equation}
we obtain:
\begin{equation}
w_{\rm \Lambda}= -\left(\frac{1}{3}+\frac{2\sqrt{\Omega_{\rm \Lambda}}}{3c}\right)
\end{equation}
Now instead if we use the Tsallis holographic energy density we can write \begin{equation*}
    \rho_{\Lambda} = \frac{3 c^2}{R_{h}^{4- 2 \sigma}} 
\end{equation*}
where $\sigma$ is the Tsallis parameter. Using this we can write $\Omega$ as \begin{equation}
    \Omega = \frac{\rho_{\Lambda}}{\rho_{cr}} = \frac{3 c^2}{R_{h}^{4-2\sigma}}
\end{equation}
One can then write (5) as \begin{equation}
    \dot{R_{h}} = R_{h}H-1=\frac{c}{\sqrt{\Omega_\Lambda} R_{h}^{1-\sigma}}-1
\end{equation}
Furthermore, the equation for $\dot{\rho}_{\Gamma} $ in this scenario takes the shape \begin{equation}
\dot{\rho}_{\Lambda} = - (4-2 \sigma) \rho H \left[ 1 - \frac{\sqrt{\Omega_{\Lambda} } R_{h}^{1-\sigma} }{c} \right]
\end{equation}
And finally this leads us to the equation of state parameter to be \begin{equation} \label{w1}
    w_{\Lambda} = - \Bigg[ \frac{2 \sigma - 1}{3} + \frac{(4-2\sigma) \sqrt{\Omega_\Lambda} R_{h}^{1-\sigma} }{3 c} \Bigg]
\end{equation}
Two things to observe here is that firstly, these equations are consistent with the usual HDE model represented by \eqref{1} as these equations reduce to that when $\sigma = 1$, as it should for any Tsallis HDE model. The second thing which is more interesting is that one observes that $w_{\Lambda} $ becomes very dynamical and acquires time dependence because of the existence of the factor $R_{h}^{1- \sigma }$ in its expression. One can work out time dependent expression for $R_{h}$ then by considering different ansatz for both the scale factor and the Hubble parameter, see\cite{Pasqua:2014oya} where 3 different ansatz for the scale factor were used to study an HDE model in a Gauss-Bonnet cosmology. What is immediately clear in this regard in the context of Tsallis cosmology is that it provides for a far more dynamical view with regards to the phantm divide than usual HDE models and that one could cross the phantom divide even for a non interacting model, as in this case we haven't assumed any interaction between dark energy and dark matter. Furthermore, the time dependent behaviour of the w parmeter for Tsallis models can have even more interesting implications for astronomical tensions like the H0 tension possibly.
\\
\\
Now, we select the scale factor to be of the form \eqref{scalefac} \footnote{While one can certainly think about other alternative forms of the scale factor ansatz like in \cite{pasqua2014study,Pasqua:2014oya,Chattopadhyay:2013mta,Chattopadhyay:2014yda}, the different scale factor choices are quite similar in the sense that they in some way or the other represent power law forms. Hence even though we can perform the analysis for another choice of the scale factor, for now we will be focused on this form only here. }   \begin{equation*}
    a(t) = a_{0} (t_{s} - t )^n
\end{equation*}
Using this, one can find the future event horizon by solving \begin{equation}
\dot{R_{h}} = R_{h}H-1=\frac{c}{\sqrt{\Omega_\Lambda}}-1
\end{equation}
to be \begin{equation}
    R_{h} (t) = \frac{C_{1} (n-1) (t-t_{s})^n+t-t_{s} }{n-1} 
\end{equation}
where $C_{1} $ is some constant of integration. Now using this we can write \eqref{w1} as \begin{equation}
w = \frac{1}{3} \left(-\frac{\sqrt{\Omega} (4-2 \sigma ) \left(C_{1} (t-t_{s})^n+\frac{t-t_{s}}{n-1}\right)^{1-\sigma }}{c}-2 \sigma +1\right)     
\end{equation}
Now in order to proceed further, we need to set some values for the parameters. One good choice would be $c=0.3 , c_{1}=0.01 ,n=3 ,\Omega=0.69 ,t_{s}=0.002 $ with $\Omega = 0.69 $. Using these, we can plot the equation of state parameter for different values of $\sigma$ to be

\begin{figure}[H]
  \centering
  \includegraphics{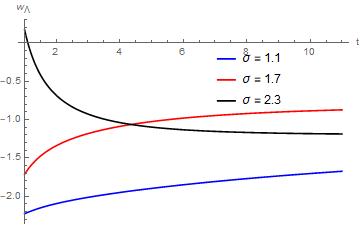}
  \caption{A plot of the dark energy EOS with time, plotted for increasing $\sigma$ values}
  \label{fig:myfigure}
\end{figure}

As is clear from Figure \ref{fig:myfigure}, one sees that while steadily increasing $\sigma$ values from over 1 to $> 2$, one sees a very curious behaviour of the EOS of the HDE. While for $\sigma =1.1$, the EOS mostly stays in the deep phantom regions, with no hope of ever even reaching near $w \to -1$, for a larger value of $\sigma = 1.7$, the EOS starts from the phantom regime, crosses the -1 barrier and ascends to some degree to the Quintessence region as well. For $\sigma > 2$ however, in this case $ \sigma = 2.3 $ for example, the EOS shows the opposite behaviour, It starts off in the deep Quintessence regime after which it crosses the -1 barrier to descend into the phantom regime to some degree. 
Furthermore, The expression for energy density in this scenario takes the form \begin{multline}
\rho(t) = \rho_{0} \exp \left(2 (\sigma -2) \left(\frac{\sqrt{o} \Gamma \left(\frac{n \sigma -1}{n-1}\right) \left(\frac{(t-t_{s})^{1-n}}{c_{1} (n-1)}+1\right)^{\sigma } \left(c_{1} (t-t_{s})^n+\frac{t-t_{s}}{n-1}\right)^{-\sigma }}{(n-1)^2 (c-c \sigma )}\right) \right. \\
+ \left. \frac{\left(c_{1} (n-1)^2 (t-t_{s})^n , 2\tilde{F}1\left(\frac{n (\sigma -1)}{n-1},\sigma ;\frac{n \sigma -1}{n-1};\frac{(t-t{s})^{1-n}}{c{1}-c_{1} n}\right)+n (\sigma -1) (t-t_{s}) , 2\tilde{F}1\left(\sigma ,\frac{n \sigma -1}{n-1};\frac{\sigma n+n-2}{n-1};\frac{(t-t{s})^{1-n}}{c{1}-c_{1} n}\right)\right)}{(n-1)^2 (c-c \sigma )} \right. \\
- \left. n \log (t-t_{s}) \right) t_{s}
\end{multline} where $2 \Tilde{F} 1$ refers to the Hypergeometric function. While the expression above looks quite humongous, which it indeed is, we can plot it for similar choice of parameters as we did before to have 
\begin{figure}[H]
  \centering
  \includegraphics{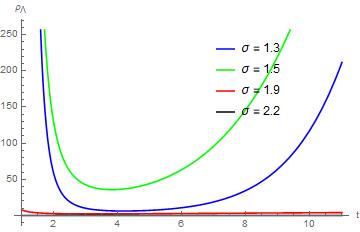}
  \caption{A plot of the DE energy density $\rho_{\Lambda} $ with time, plotted for increasing $\sigma$ values}
  \label{fig:myfigure2}
\end{figure}
One sees from figure \ref{fig:myfigure2} that for progressively increasing values of $\sigma$, the variation of the energy density with respect to time becomes smaller and smaller as for $\sigma = 1.3 $ and $\sigma = 1.5$, one sees that the energy density fluctuates a lot while for higher values of $\sigma =1.9$ and $\sigma = 2.2$, it almost takes the shape of a constant curve. 
\\
\\
This brings us to another interesting issue with usual HDE models, which the Tsallis model with event Horizon cutoff can possibly handle optimistically in in the way that we have dealt with it here. This concerns the issue of classical instabilities of perturbations in HDE models, which was firstly shown in \cite{Myung:2007pn}. The key for this lies in the squared speed of sound, which for our case would take the form \begin{equation}
    v_{s}^2 = \frac{\dot{p}}{\dot{\rho}}
\end{equation}
The sign of $v_{s}^2$ is crucial for determining the stability of background evolution, with negative $v_{s}^2$ for a model highlighting that the model has classical instabilities for any given perturbation.  In \cite{Myung:2007pn} it was shown that conventional HDE models would have this instability and recently there was one paper which also showed that Tsallis HDE can bypass this issue in interacting scenarios \cite{bhattacharjee2020interacting,zubair2022thermodynamics} but nobody has considered whether these issues can be bypassed in a non interacting scenario and so here we can actually show that the same can happen for a non interacting scenario as well. The squared sound speed for our model can be written as \begin{equation}
   v_{sv}^2 = \frac{1}{3} \left(\frac{\sqrt{\Omega} (\sigma -1) R_{h}(t) R'_{h}(t)}{h(t) \left(c R_{h}(t)^{\sigma}-\sqrt{\Omega} R_{h}(t)\right)}+\frac{2 \sqrt{\Omega} (\sigma -2) R_{h}(t)^{1-\sigma}}{c}-2 \sigma +1\right)    
\end{equation}
Plugging in the expression for $R_{h} (t) $ from before, we get \begin{multline}
    v_{sv}^2 = \frac{1}{3} \left(-\frac{\sqrt{\Omega} (1-\sigma) \left(c_{1} (n-1) n (t-t_{s})^{n-1}+1\right) \left(c_{1} (t-t_{s})^n+\frac{t-t_{s}}{n-1}\right)^{1-\sigma}}{(n-1) h(t) \left(c-\sqrt{\Omega} \left(c_{1} (t-t_{s})^n+\frac{t-t_{s}}{n-1}\right)^{1-\sigma}\right)} \right.\\
    \left.- \frac{\sqrt{\Omega} (4-2 \sigma) \left(c_{1} (t-t_{s})^n+\frac{t-t_{s}}{n-1}\right)^{1-\sigma}}{c}-2 \sigma +1\right)
\end{multline}
This equation can be plotted as \begin{figure}[H]
  \centering
  \includegraphics{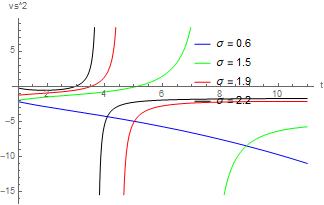}
  \caption{A plot of the squared sound speed $v_{s}^2 $ for the event horizon cutoff Tsallis model with time, plotted for increasing $\sigma$ values}
  \label{fig:myfigure3}
\end{figure} 
One can clearly see from figure \ref{fig:myfigure3} that one can get rid of the stability issues persistent in normal HDE models in a Tsallis HDE scenario. Particularly, one sees that for very small values of $\sigma$, in this $\sigma = 0.6$ (which is even lower than the $\sigma$ for recovering simple HDE ), there is no hope of escaping the instability issues as $v_{s}^2$ is always negative. However, as one increases the values of $\sigma$ and one gradually starts to deviate away from usual HDE models, there is a gradual departure from the instability problem $v_{s}^2$ for all of $\sigma = (1.5,1.9,2.2) $ obtains positive values even though they could initially have negative values. This shows that Tsallis models with higher values of the tsallis parameters offer classical stability and as one deviates away from the usual HDE scenario, one obtains models which have more and more classical stability in the face of perturbations.But even the higher values of $\sigma$ eventually tend to instability as time progresses on and on. It is helpful to see the behaviour of the squared sound speed values at artbitrarily large values \begin{figure}[H]
  \centering
  \includegraphics{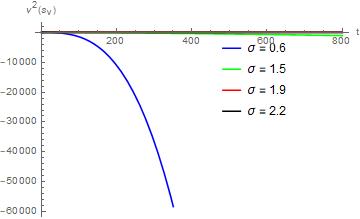}
  \caption{A plot of the squared sound speed $v_{s}^2 $ for the event horizon cutoff viscous Tsallis model with time, plotted for increasing $\sigma$ values in an arbitrarily large timeframe}
  \label{fig:myfigure7}
\end{figure} 
One sees from figure \ref{fig:myfigure7} that far ahead in time, the models corresponding to all these various values of $\sigma$ eventually acquire classical instability as they have negative squared sound speeds. 
\\
\\
\subsection{Viscous fluid case}
While we are already obtaining very interesting results for the Tsallis model for conventional fluids with $p = w \rho$, one can go beyond this too. Particularly, let's consider the case of viscous and Chaplygin gas fluids. The equation of state that we consider for the viscous fluid configuration is given by \cite{Nojiri:2006zh,Timoshkin:2023pfb} \begin{equation}
    p = w \rho - 3 \epsilon_{0} H
\end{equation}
where $\epsilon_{0} $ is a thermodynamic parameter which can be considered to be either time dependent or time independent. As we shall see, this parameter does not directly impact any of our calculations. Using \eqref{1}, we can write the continuity equation in this case to be \begin{equation}
    - (4-2 \sigma) \rho H \left[ 1 - \frac{\sqrt{\Omega_{\Lambda} } R_{h}^{1-\sigma} }{c} \right] + 3 H( \rho (1 + w_{v}) - 3 \epsilon_{0} H) = 0 
\end{equation}
where $w_{v} $ is the EOS for HDE in the case of the viscous fluid configuration. Working with the above equation, after some effort one can reach for w to be \begin{equation}
    w_{v} = w_{v} = -\frac{1}{3} (4-2 \sigma) \left(1-\frac{\sqrt{\Omega} R_{h}(t)^{1-\sigma}}{c}\right)-1
\end{equation}
Using the expression for $R_{h}(t)$ one can then write \begin{equation}
   w_{v} = -\frac{1}{3} (4-2 \sigma ) \left(1-\frac{\sqrt{o} \left(c_{1} (t-t_{s})^n+\frac{t-t_{s}}{n-1}\right)^{1-\sigma }}{c}\right)-1 
\end{equation}
This equation can be plotted to be \begin{figure}[H]
  \centering
  \includegraphics{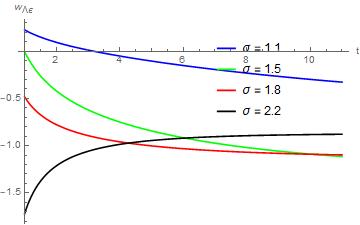}
  \caption{A plot of the dark energy EOS for the event horizon cutoff Tsallis model with viscous fluid configuration $w_{v} $ with time, plotted for increasing $\sigma$ values}
  \label{fig:myfigure4}
\end{figure} 
As one sees from figure \ref{fig:myfigure4}, the dark energy EOS again crosses the phantom divide but the behaviour here is completely the opposite to what one observes for the non-viscous fluid case in figure \ref{fig:myfigure}. Particularly, one sees that for smaller values of $\sigma$, like $\sigma =1.1$ in this case, which represents possibly the smallest deviation from the normal HDE to the Tsallis model,the EOS starts off in the deep Quintessence region and never really hits the $w=-1$ region, let alone crossing the phantom barrier. While the higher values of $\sigma$ in this case like $\sigma=(1.5,1.8) $ start off in the Quintessence region and descend to Phantom while the very high values like $\sigma =2.2$, start from the deep phantom region and only go over ever so slightly above -1 into a very small patch of the Quintessence regime. The phantom divide happens again but in a completely opposite fashion to the trend seen in the non-viscous fluid case. The energy density of dark energy fluctuates in pretty much the same way as in figure \ref{fig:myfigure2} shown for the non-viscous regime as it does in the viscous regime and so we dont need to show its plot here. 
\\
\\
The sound speed squared in this case turns out to be \begin{equation}
    v_{sv}^2 = \frac{1}{3} \left(\sqrt{\Omega} \left(\frac{(\sigma -1) \dot{R_{h}(t)}}{h(t) \left(\sqrt{\Omega} R_{h}(t)-c R_{h}(t)^{\sigma}\right)}+\frac{2 (\sigma -2) R_{h}(t)^{1-\sigma}}{c}\right)-2 \sigma +1\right) 
\end{equation}
which using the expression we have for the future event horizon becomes \begin{multline}
  v_{sv}^2 = \frac{1}{3} (4-2 \sigma) \left(-\frac{\sqrt{\Omega} \left(c_{1} (t-t_{s})^n+\frac{t-t_{s}}{n-1}\right)^{1-\sigma}}{c} \right. \\
  + \left. \frac{\sqrt{\Omega} (\sigma -1) \left(c_{1} (n-1) n (t-t_{s})^n+t-t_{s}\right)}{2 (\sigma -2) \left(n \sqrt{\Omega} \left(c_{1} (n-1) (t-t_{s})^n+t-t_{s}\right)-c (n-1) n \left(c_{1} (t-t_{s})^n+\frac{t-t_{s}}{n-1}\right)^{\sigma}\right)}+1\right)-1   
\end{multline}
This can be plotted against time as \begin{figure}[H]
  \centering
  \includegraphics{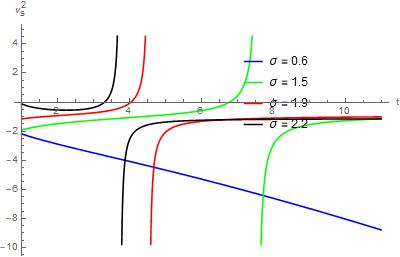}
  \caption{A plot of the squared sound speed $v_{s}^2 $ for the event horizon cutoff viscous Tsallis model with time, plotted for increasing $\sigma$ values}
  \label{fig:myfigure5}
\end{figure} 
One clearly sees that there is a similarity to the trend in the graphs of the squared sound speed in both the viscous and non-viscous cases given by figure \ref{fig:myfigure3} and figure \ref{fig:myfigure5} respectively. We again see that for very small values of $\sigma$ there is no hope to avoiding classical instabilities even in the viscous Tsallis model while one can hope for better results as one increases the value of $\sigma$. Hence, one sees that as one increases the deviation from the usual HDE model and with the effects of the Tsallis model becoming more and more apparent, the stability of the model has a hope of becoming stable. But again, even the higher values of $\sigma$ eventually tend to instability. In this scenario it is especially helpful to see the behaviour at arbitrarily large times, the plot for which is \begin{figure}[H]
  \centering
  \includegraphics{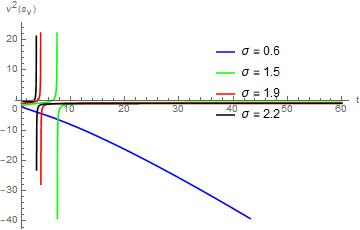}
  \caption{A plot of the squared sound speed $v_{s}^2 $ for the event horizon cutoff viscous Tsallis model with time, plotted for increasing $\sigma$ values at arbitrarily large times}
  \label{fig:myfigure6}
\end{figure} 
As it is clear from figure \ref{fig:myfigure6}, as time progresses to very large values, models for all values of $\sigma$ eventually attain negative squared sound speed values. Hence all these models likely become unstable for perturbations in large timeframes. 
\\
\\
\subsection{Chaplygin gas case}
Till now both the models that we have discussed have given interesting results from the perspective of the phantom divide, showing that the Tsallis model can show such divide for nice range of the Tsallis parameter for both viscous and non viscous fluid configurations. But although even in these paradigms we are seeing small hints of escaping the issues of classical instability as pointed out in \cite{Myung:2007pn}, the instability eventually takes over in these models as time progresses. Now we turn our attention to the generalized Chaplygin gas model, characterized by the EOS \begin{equation}
    p = - \frac{A}{\rho^{\alpha}}
\end{equation}
where A and $\alpha$ are assumed to be positive constants, where for $\alpha = 1$, one obtains the usual Chaplygin gas model. Now for a Tsallis HDe scenario with a Chaplygin gas configuration, we can write the continuinty equation to be \begin{equation}
    - (4-2 \sigma) \rho H \left[ 1 - \frac{\sqrt{\Omega_{\Lambda} } R_{h}^{1-\sigma} }{c} \right] + 3 H( \rho - \frac{A}{\rho^{\alpha}} ) = 0
\end{equation}
from which we can write the EOS to be \begin{equation}
    w_{c} = +\frac{1}{3} (4-2 \sigma) \left(1-\frac{\sqrt{\Omega} R_{h}(t)^{1-\sigma}}{c}\right)-1 
\end{equation}
Using the expression for the future event horizon, we can write \begin{equation}
   w_{c} =  \frac{1}{3} (4-2 \sigma ) \left(1-\frac{\sqrt{o} \left(c_{1} (t-t_{s})^n+\frac{t-t_{s}}{n-1}\right)^{1-\sigma }}{c}\right)-1
\end{equation}
This can be plotted as follows\begin{figure}[H]
  \centering
  \includegraphics{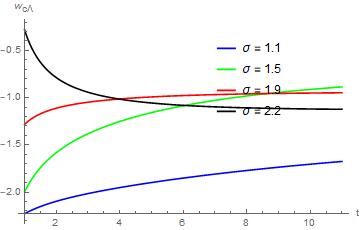}
  \caption{A plot of the dark energy EOS for the event horizon cutoff Tsallis model with Chaplygin gas configuration $w_{c} $ with time, plotted for increasing $\sigma$ values}
  \label{fig:myfigure8}
\end{figure} 
As one can see from figure \ref{fig:myfigure8}, one sees that for relatively smaller values of time the larger values of $\sigma$ (in this case $\sigma = (1.5,1.9,2.2) $ cross the -1 barrier starting either from the quintessence or phantom regime while the smallest value of $\sigma$ which in this case is $\sigma =1.1$, appears to not be able to do so in this timeframe. Plotting for arbitrarily large times, however, gives us the following plot \begin{figure}[H]
  \centering
  \includegraphics{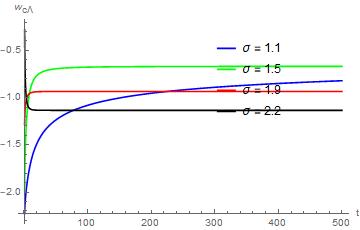}
  \caption{A plot of the dark energy EOS for the event horizon cutoff Tsallis model with Chaplygin gas configuration $w_{c} $ with time, plotted for increasing $\sigma$ values, plotted for a large timeframe}
  \label{fig:myfigure9}
\end{figure} 
For large values of time, or equivalently speaking as time progresses further and further, one sees from figure \ref{fig:myfigure9} that even the model with the slightest deviation from the simple HDE, $\sigma =1.1$, also crosses the -1 barrier and eventually ascends into the Quintessence regime. While the model for $\sigma = 1.5$ becomes well settled in the Quintessence regime with $\sigma =(1.9,2.2) $ models becoming settled in the phantom regime. The thing to note, very interestingly, is that the Chaplygin gas parameters, A and $\alpha$, do not impact the behaviour of the equation of state at all and do not factor into the equation in any case. One can also consider the energy density of the HDE in this case, which can be written as \begin{equation}
    \rho_{\Lambda c} = 3^{\frac{1}{\alpha +1}} \left(\frac{a c R_{h}(t)^{\sigma}}{c (2 \sigma -1) R_{h}(t)^{\sigma}-2 \sqrt{\Omega} (\sigma -2) R_{h}(t)}\right)^{\frac{1}{\alpha +1}} 
\end{equation}
Plotting the expression of the future event horizon we get \begin{equation}
   \rho_{\Lambda c} = 3^{\frac{1}{\alpha+1}} \left(\frac{a}{3-(4-2 \sigma ) \left(1-\frac{\sqrt{\Omega} \left(c_{1} (t-t_{s})^n+\frac{t-t_{s}}{n-1}\right)^{1-\sigma }}{c}\right)}\right)^{\frac{1}{\alpha+1}}
\end{equation}
And plotting this for different values of A, we can get \begin{figure}[H]
  \centering
  \includegraphics[width=1\textwidth]{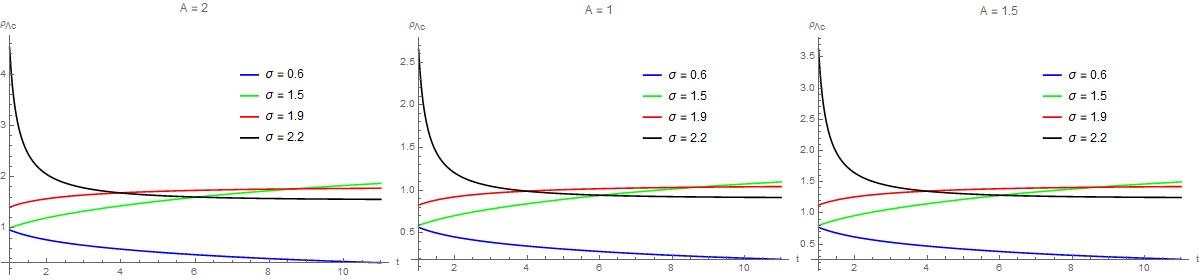}
  \caption{A plot of the DE energy density $\rho_{\Lambda c}$, plotted for increasing $\sigma$ values plotted for different values of A against time}
  \label{fig:myfigure10}
\end{figure}
As one sees from figure \ref{fig:myfigure10} that the density fluctuates in different ways for all values of $\sigma$ but the trends of the evolution of the densities stays quite the same for different values of A. Plotting for one of these values at large times, say $A=2$, we have \begin{figure}[H]
  \centering
  \includegraphics[width=0.8\textwidth]{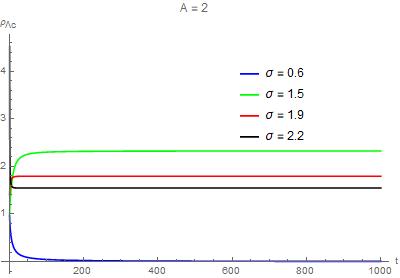}
  \caption{A plot of the DE energy density $\rho_{\Lambda c}$, plotted for increasing $\sigma$ values plotted for different values of A against time}
  \label{fig:myfigure11}
\end{figure}
We see from figure \ref{fig:myfigure11} that eventually its the value which is not too high like $\sigma = 2.2$ or not too low like $\sigma =1.1$m but rather quite in the intermediate which is $\sigma = 1.5$ which eventually has the highest values of the energy density of the HDE. The evolution of all the densities in different models is again different but is pretty much a smooth transition from what one observed for smaller timeframes in figure \ref{fig:myfigure10}.
\\
\\
Now we finally turn our attention to what perhaps is probably the most surprising discovery here. The squared speed in the case of the Chaplygin gas scenario can be written as \begin{equation}
v_{sc}^2 = \frac{1}{3} \alpha \left(3-(4-2 \sigma) \left(1-\frac{\sqrt{\Omega} R_{h}(t)^{1-\sigma}}{c}\right)\right)     
\end{equation}
And again by using the expression of the future event horizon we can write \begin{equation}
v_{sc}^2 = \alpha-\frac{1}{3} \alpha (4-2 \sigma ) \left(1-\frac{\sqrt{\Omega} \left(c_{1} (t-t_{s})^n+\frac{t-t_{s}}{n-1}\right)^{1-\sigma }}{c}\right)    
\end{equation}
Plotting this against time, in figure \ref{fig13} we plot the sound speed squared for different values of $\sigma$ setting $\alpha = 0.3$ (the trend of the sound speed stays the same for $<\alpha \leq 1$) \begin{figure}[H]
  \centering
  \includegraphics[width=0.8\textwidth]{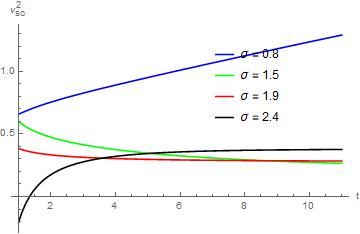}
  \caption{A plot of the squared sound speed $v_{sc}^2$, plotted for increasing $\sigma$ values plotted for $\alpha=0.3$ against time}
  \label{fig13}
\end{figure}
The plot which is being seen in figure \ref{fig13} is quite interesting. One sees that for all values of $\sigma$ in this case, which are $\sigma = (0.8,1.5,1.9,2.4)$, there is a clear and very evident preference of the squared sound speed towards highly positive values. In fact, only the highest value of $\sigma$ in this case which is $\sigma = 2.4$, picks up negative values of the squared sound speed in the very beginning while it very quickly ascends to positive values. In a larger timeframe, the plot takes the shape \begin{figure}[H]
  \centering
  \includegraphics[width=0.8\textwidth]{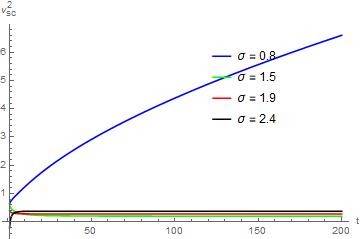}
  \caption{A plot of the squared sound speed $v_{sc}^2$, plotted for increasing $\sigma$ values plotted for $\alpha=0.3$ against time}
  \label{fig14}
\end{figure} 
One sees from figure \ref{fig14} that in arbitrarily large time frames or equivalently speaking, as time progresses further and further on, one sees that various $\sigma$ models very easily and definitively maintain classical stability as they eventually have quite substantial positive values for the squared sound speed. This definitively stable behaviour from the perturbations point of view is seen very convincingly in the case of the Tsallis Chaplygin gas scenario, something which we only saw in very small patches for the previous two Tsallis models that we considered.
\\
\\
\section{Conclusions}
In this work we considered the Tsallis Holographic dark energy model in a scale factor ansatz based approach. We considered the event horizon cut-off for our analysis while considering three different cosmological settings for our model namely with non-viscous, viscous and Chaplygin gas configurations. We were able to show that in all such cases, one can clearly see that the Tsallis models cross the phantom divide but stability issues persist throughout the reasonable range of the Tsallis parameter in both viscous and non-viscous cases. But interestingly, the Chaplygin gas models present us with more stable dark energy scenarios as they they give long term stability. 

\section*{Acknowledgements}
The research by M.K. was carried out at Southern Federal University with financial support from the Ministry of Science and Higher Education of the Russian Federation (State contract GZ0110/23-10-IF). The work of A.T. was supported by Russian Foundation for Basic Research; Project No. 20-52-05009. 

\bibliography{Jcitations}

\begin{thebibliography}{10}

\bibitem{SupernovaSearchTeam:1998fmf}
Adam~G. Riess et~al.
\newblock {Observational evidence from supernovae for an accelerating universe and a cosmological constant}.
\newblock {\em Astron. J.}, 116:1009--1038, 1998.

\bibitem{Weinberg:1988cp}
Steven Weinberg.
\newblock {The Cosmological Constant Problem}.
\newblock {\em Rev. Mod. Phys.}, 61:1--23, 1989.

\bibitem{Lombriser:2019jia}
Lucas Lombriser.
\newblock {On the cosmological constant problem}.
\newblock {\em Phys. Lett. B}, 797:134804, 2019.

\bibitem{Padmanabhan:2002ji}
T.~Padmanabhan.
\newblock {Cosmological constant: The Weight of the vacuum}.
\newblock {\em Phys. Rept.}, 380:235--320, 2003.

\bibitem{Capozziello:2011et}
Salvatore Capozziello and Mariafelicia De~Laurentis.
\newblock {Extended Theories of Gravity}.
\newblock {\em Phys. Rept.}, 509:167--321, 2011.

\bibitem{Nojiri:2010wj}
Shin'ichi Nojiri and Sergei~D. Odintsov.
\newblock {Unified cosmic history in modified gravity: from F(R) theory to Lorentz non-invariant models}.
\newblock {\em Phys. Rept.}, 505:59--144, 2011.

\bibitem{Nojiri:2017ncd}
S.~Nojiri, S.~D. Odintsov, and V.~K. Oikonomou.
\newblock {Modified Gravity Theories on a Nutshell: Inflation, Bounce and Late-time Evolution}.
\newblock {\em Phys. Rept.}, 692:1--104, 2017.

\bibitem{Zlatev:1998tr}
Ivaylo Zlatev, Li-Min Wang, and Paul~J. Steinhardt.
\newblock {Quintessence, cosmic coincidence, and the cosmological constant}.
\newblock {\em Phys. Rev. Lett.}, 82:896--899, 1999.

\bibitem{Tsujikawa:2013fta}
Shinji Tsujikawa.
\newblock {Quintessence: A Review}.
\newblock {\em Class. Quant. Grav.}, 30:214003, 2013.

\bibitem{Faraoni:2000wk}
Valerio Faraoni.
\newblock {Inflation and quintessence with nonminimal coupling}.
\newblock {\em Phys. Rev. D}, 62:023504, 2000.

\bibitem{Gasperini:2001pc}
M.~Gasperini, F.~Piazza, and G.~Veneziano.
\newblock {Quintessence as a runaway dilaton}.
\newblock {\em Phys. Rev. D}, 65:023508, 2002.

\bibitem{Capozziello:2003tk}
Salvatore Capozziello, Sante Carloni, and Antonio Troisi.
\newblock {Quintessence without scalar fields}.
\newblock {\em Recent Res. Dev. Astron. Astrophys.}, 1:625, 2003.

\bibitem{Capozziello:2002rd}
Salvatore Capozziello.
\newblock {Curvature quintessence}.
\newblock {\em Int. J. Mod. Phys. D}, 11:483--492, 2002.

\bibitem{Odintsov:2023weg}
S.~D. Odintsov, V.~K. Oikonomou, I.~Giannakoudi, F.~P. Fronimos, and E.~C. Lymperiadou.
\newblock {Recent Advances on Inflation}.
\newblock {\em Symmetry}, 15:9, 2023.

\bibitem{Sahni:2002dx}
Varun Sahni and Yuri Shtanov.
\newblock {Brane world models of dark energy}.
\newblock {\em JCAP}, 11:014, 2003.

\bibitem{Sami:2004xk}
M.~Sami and V.~Sahni.
\newblock {Quintessential inflation on the brane and the relic gravity wave background}.
\newblock {\em Phys. Rev. D}, 70:083513, 2004.

\bibitem{Tretyakov:2005en}
Petr Tretyakov, Aleksey Toporensky, Yuri Shtanov, and Varun Sahni.
\newblock {Quantum effects, soft singularities and the fate of the universe in a braneworld cosmology}.
\newblock {\em Class. Quant. Grav.}, 23:3259--3274, 2006.

\bibitem{Chen:2008ca}
Songbai Chen, Bin Wang, and Jiliang Jing.
\newblock {Dynamics of interacting dark energy model in Einstein and Loop Quantum Cosmology}.
\newblock {\em Phys. Rev. D}, 78:123503, 2008.

\bibitem{Fu:2008gh}
Xiangyun Fu, Hong~Wei Yu, and Puxun Wu.
\newblock {Dynamics of interacting phantom scalar field dark energy in Loop Quantum Cosmology}.
\newblock {\em Phys. Rev. D}, 78:063001, 2008.

\bibitem{Bonanno:2001hi}
A.~Bonanno and M.~Reuter.
\newblock {Cosmology with selfadjusting vacuum energy density from a renormalization group fixed point}.
\newblock {\em Phys. Lett. B}, 527:9--17, 2002.

\bibitem{Bonanno:2001xi}
A.~Bonanno and M.~Reuter.
\newblock {Cosmology of the Planck era from a renormalization group for quantum gravity}.
\newblock {\em Phys. Rev. D}, 65:043508, 2002.

\bibitem{Bentivegna:2003rr}
Eloisa Bentivegna, Alfio Bonanno, and Martin Reuter.
\newblock {Confronting the IR fixed point cosmology with high redshift supernova data}.
\newblock {\em JCAP}, 01:001, 2004.

\bibitem{Reuter:2005kb}
M.~Reuter and Frank Saueressig.
\newblock {From big bang to asymptotic de Sitter: Complete cosmologies in a quantum gravity framework}.
\newblock {\em JCAP}, 09:012, 2005.

\bibitem{Bonanno:2007wg}
Alfio Bonanno and Martin Reuter.
\newblock {Entropy signature of the running cosmological constant}.
\newblock {\em JCAP}, 08:024, 2007.

\bibitem{Weinberg:2009wa}
Steven Weinberg.
\newblock {Asymptotically Safe Inflation}.
\newblock {\em Phys. Rev. D}, 81:083535, 2010.

\bibitem{Trivedi:2022svr}
Oem Trivedi and Maxim Khlopov.
\newblock {Singularity formation in asymptotically safe cosmology with inhomogeneous equation of state}.
\newblock {\em JCAP}, 11:007, 2022.

\bibitem{Trivedi:2022ipa}
Oem Trivedi and Maxim Khlopov.
\newblock {On finite time singularities in scalar field dark energy models based in the RS-II Braneworld}.
\newblock {\em Eur. Phys. J. C}, 82(9):800, 2022.

\bibitem{Planck:2018vyg}
N.~Aghanim et~al.
\newblock {Planck 2018 results. VI. Cosmological parameters}.
\newblock {\em Astron. Astrophys.}, 641:A6, 2020.
\newblock [Erratum: Astron.Astrophys. 652, C4 (2021)].

\bibitem{riess2019large}
Adam~G Riess, Stefano Casertano, Wenlong Yuan, Lucas~M Macri, and Dan Scolnic.
\newblock Large magellanic cloud cepheid standards provide a 1\% foundation for the determination of the hubble constant and stronger evidence for physics beyond $\lambda$cdm.
\newblock {\em The Astrophysical Journal}, 876(1):85, 2019.

\bibitem{riess2021comprehensive}
Adam~G Riess, Wenlong Yuan, Lucas~M Macri, Dan Scolnic, Dillon Brout, Stefano Casertano, David~O Jones, Yukei Murakami, Louise Breuval, Thomas~G Brink, et~al.
\newblock A comprehensive measurement of the local value of the hubble constant with 1 km/s/mpc uncertainty from the hubble space telescope and the sh0es team.
\newblock {\em arXiv preprint arXiv:2112.04510}, 2021.

\bibitem{tHooft:1993dmi}
Gerard 't~Hooft.
\newblock {Dimensional reduction in quantum gravity}.
\newblock {\em Conf. Proc. C}, 930308:284--296, 1993.

\bibitem{Susskind:1994vu}
Leonard Susskind.
\newblock {The World as a hologram}.
\newblock {\em J. Math. Phys.}, 36:6377--6396, 1995.

\bibitem{Bousso:1999xy}
Raphael Bousso.
\newblock {A Covariant entropy conjecture}.
\newblock {\em JHEP}, 07:004, 1999.

\bibitem{Cohen:1998zx}
Andrew~G. Cohen, David~B. Kaplan, and Ann~E. Nelson.
\newblock {Effective field theory, black holes, and the cosmological constant}.
\newblock {\em Phys. Rev. Lett.}, 82:4971--4974, 1999.

\bibitem{Wang:2016och}
Shuang Wang, Yi~Wang, and Miao Li.
\newblock {Holographic Dark Energy}.
\newblock {\em Phys. Rept.}, 696:1--57, 2017.

\bibitem{Nojiri:2005pu}
Shin'ichi Nojiri and Sergei~D. Odintsov.
\newblock {Unifying phantom inflation with late-time acceleration: Scalar phantom-non-phantom transition model and generalized holographic dark energy}.
\newblock {\em Gen. Rel. Grav.}, 38:1285--1304, 2006.

\bibitem{Nojiri:2017opc}
Shin'ichi Nojiri and S.~D. Odintsov.
\newblock {Covariant Generalized Holographic Dark Energy and Accelerating Universe}.
\newblock {\em Eur. Phys. J. C}, 77(8):528, 2017.

\bibitem{Nojiri:2021iko}
Shin'ichi Nojiri, Sergei~D. Odintsov, and Tanmoy Paul.
\newblock {Different Faces of Generalized Holographic Dark Energy}.
\newblock {\em Symmetry}, 13(6):928, 2021.

\bibitem{Nojiri:2020wmh}
Shinichi Nojiri, S.~D. Odintsov, V.~K. Oikonomou, and Tanmoy Paul.
\newblock {Unifying Holographic Inflation with Holographic Dark Energy: a Covariant Approach}.
\newblock {\em Phys. Rev. D}, 102(2):023540, 2020.

\bibitem{Oliveros:2022biu}
A.~Oliveros, M.~A. Sabogal, and Mario~A. Acero.
\newblock {Barrow holographic dark energy with Granda\textendash{}Oliveros cutoff}.
\newblock {\em Eur. Phys. J. Plus}, 137(7):783, 2022.

\bibitem{Granda:2008dk}
L.~N. Granda and A.~Oliveros.
\newblock {Infrared cut-off proposal for the Holographic density}.
\newblock {\em Phys. Lett. B}, 669:275--277, 2008.

\bibitem{Khurshudyan:2016gmb}
Martiros Khurshudyan.
\newblock {Viscous holographic dark energy universe with Nojiri-Odintsov cut-off}.
\newblock {\em Astrophys. Space Sci.}, 361(12):392, 2016.

\bibitem{Khurshudyan:2016uql}
Martiros Khurshudyan.
\newblock {On a holographic dark energy model with a Nojiri-Odintsov cut-off in general relativity}.
\newblock {\em Astrophys. Space Sci.}, 361(7):232, 2016.

\bibitem{Belkacemi:2011zk}
Moulay-Hicham Belkacemi, Mariam Bouhmadi-Lopez, Ahmed Errahmani, and Taoufiq Ouali.
\newblock {The holographic induced gravity model with a Ricci dark energy: smoothing the little rip and big rip through Gauss-Bonnet effects?}
\newblock {\em Phys. Rev. D}, 85:083503, 2012.

\bibitem{Zhang:2011zze}
Linsen Zhang, Puxun Wu, and Hongwei Yu.
\newblock {Unifying dark energy and dark matter with the modified Ricci model}.
\newblock {\em Eur. Phys. J. C}, 71:1588, 2011.

\bibitem{Setare:2010zy}
M.~R. Setare and Mubasher Jamil.
\newblock {Statefinder diagnostic and stability of modified gravity consistent with holographic and new agegraphic dark energy}.
\newblock {\em Gen. Rel. Grav.}, 43:293--303, 2011.

\bibitem{Nozari:2009zk}
Kourosh Nozari and Narges Rashidi.
\newblock {Holographic Dark Energy from a Modified GBIG Scenario}.
\newblock {\em Int. J. Mod. Phys. D}, 19:219--231, 2010.

\bibitem{Sheykhi:2009dz}
Ahmad Sheykhi.
\newblock {Interacting holographic dark energy in Brans-Dicke theory}.
\newblock {\em Phys. Lett. B}, 681:205--209, 2009.

\bibitem{Xu:2009xi}
Lixin Xu, Jianbo Lu, and Wenbo Li.
\newblock {Generalized Holographic and Ricci Dark Energy Models}.
\newblock {\em Eur. Phys. J. C}, 64:89--95, 2009.

\bibitem{Wei:2009au}
Hao Wei.
\newblock {Modified Holographic Dark Energy}.
\newblock {\em Nucl. Phys. B}, 819:210--224, 2009.

\bibitem{Setare:2008hm}
M.~R. Setare and E.~N. Saridakis.
\newblock {Correspondence between Holographic and Gauss-Bonnet dark energy models}.
\newblock {\em Phys. Lett. B}, 670:1--4, 2008.

\bibitem{Saridakis:2007wx}
E.~N. Saridakis.
\newblock {Holographic Dark Energy in Braneworld Models with a Gauss-Bonnet Term in the Bulk. Interacting Behavior and the w =-1 Crossing}.
\newblock {\em Phys. Lett. B}, 661:335--341, 2008.

\bibitem{Setare:2006yj}
M~R Setare.
\newblock {The Holographic dark energy in non-flat Brans-Dicke cosmology}.
\newblock {\em Phys. Lett. B}, 644:99--103, 2007.

\bibitem{Felegary:2016znh}
F.~Felegary, F.~Darabi, and M.~R. Setare.
\newblock {Interacting holographic dark energy model in Brans\textendash{}Dicke cosmology and coincidence problem}.
\newblock {\em Int. J. Mod. Phys. D}, 27(03):1850017, 2017.

\bibitem{Dheepika:2021fqv}
M.~Dheepika and Titus~K. Mathew.
\newblock {Tsallis holographic dark energy reconsidered}.
\newblock {\em Eur. Phys. J. C}, 82(5):399, 2022.

\bibitem{Brevik:2022aqp}
I.~Brevik and A.~V. Timoshkin.
\newblock {Holographic representation of the unified early and late universe via a viscous dark fluid}.
\newblock {\em Int. J. Geom. Meth. Mod. Phys.}, 19(08):2250113, 2022.

\bibitem{Brevik:2023vaf}
I.~Brevik and A.~V. Timoshkin.
\newblock {Holographic description of the dissipative unified dark fluid model with axion field}.
\newblock {\em Int. J. Mod. Phys. D}, 32(13):2350085, 2023.

\bibitem{Nojiri:2022aof}
Shin'ichi Nojiri, Sergei~D. Odintsov, and Valerio Faraoni.
\newblock {From nonextensive statistics and black hole entropy to the holographic dark universe}.
\newblock {\em Phys. Rev. D}, 105(4):044042, 2022.

\bibitem{Nojiri:2019skr}
Shin'ichi Nojiri, Sergei~D. Odintsov, and Emmanuel~N. Saridakis.
\newblock {Modified cosmology from extended entropy with varying exponent}.
\newblock {\em Eur. Phys. J. C}, 79(3):242, 2019.

\bibitem{Pasqua:2014oya}
Antonio Pasqua, Surajit Chattopadhyay, Martiros Khurshudyan, and Ayman~A. Aly.
\newblock {Behavior of Holographic Ricci Dark Energy in Scalar Gauss-Bonnet Gravity for Different Choices of the Scale Factor}.
\newblock {\em Int. J. Theor. Phys.}, 53(9):2988--3013, 2014.

\bibitem{Chattopadhyay:2013mta}
Surajit Chattopadhyay and Antonio Pasqua.
\newblock {Holographic DBI-Essence Dark Energy via Power-Law Solution of the Scale Factor}.
\newblock {\em Int. J. Theor. Phys.}, 52:3945--3952, 2013.

\bibitem{Chattopadhyay:2014yda}
Surajit Chattopadhyay, Antonio Pasqua, and Martiros Khurshudyan.
\newblock {New holographic reconstruction of scalar field dark energy models in the framework of chameleon Brans-Dicke cosmology}.
\newblock {\em Eur. Phys. J. C}, 74(9):3080, 2014.

\bibitem{Jawad:2016tne}
Abdul Jawad, Nadeem Azhar, and Shamaila Rani.
\newblock {Entropy corrected holographic dark energy models in modified gravity}.
\newblock {\em Int. J. Mod. Phys. D}, 26(04):1750040, 2016.

\bibitem{Tsallis:2012js}
Constantino Tsallis and Leonardo J.~L. Cirto.
\newblock {Black hole thermodynamical entropy}.
\newblock {\em Eur. Phys. J. C}, 73:2487, 2013.

\bibitem{Barboza:2014yfe}
Ed\'esio~M. Barboza, Jr., Rafael da~C. Nunes, Everton M.~C. Abreu, and Jorge Ananias~Neto.
\newblock {Dark energy models through nonextensive Tsallis\textquoteright{} statistics}.
\newblock {\em Physica A}, 436:301--310, 2015.

\bibitem{Nunes:2015xsa}
Rafael~C. Nunes, Ed\'esio~M. Barboza, Jr., Everton M.~C. Abreu, and Jorge~Ananias Neto.
\newblock {Probing the cosmological viability of non-gaussian statistics}.
\newblock {\em JCAP}, 08:051, 2016.

\bibitem{moradpour2016implications}
H~Moradpour.
\newblock Implications, consequences and interpretations of generalized entropy in the cosmological setups.
\newblock {\em International Journal of Theoretical Physics}, 55:4176--4184, 2016.

\bibitem{Tavayef:2018xwx}
M.~Tavayef, A.~Sheykhi, Kazuharu Bamba, and H.~Moradpour.
\newblock {Tsallis Holographic Dark Energy}.
\newblock {\em Phys. Lett. B}, 781:195--200, 2018.

\bibitem{zadeh2018note}
M~Abdollahi Zadeh, A~Sheykhi, H~Moradpour, and Kazuharu Bamba.
\newblock Note on tsallis holographic dark energy.
\newblock {\em The European Physical Journal C}, 78:1--11, 2018.

\bibitem{aly2019tsallis}
Ayman~A Aly et~al.
\newblock Tsallis holographic dark energy with granda-oliveros scale in ()-dimensional frw universe.
\newblock {\em Advances in Astronomy}, 2019, 2019.

\bibitem{srivastava2020tsallis}
Vandna Srivastava and Umesh~Kumar Sharma.
\newblock Tsallis holographic dark energy with hybrid expansion law.
\newblock {\em International Journal of Geometric Methods in Modern Physics}, 17(11):2050144, 2020.

\bibitem{sharma2021tsallis}
Umesh~Kumar Sharma and Vandna Srivastava.
\newblock Tsallis hde with an ir cutoff as ricci horizon in a flat flrw universe.
\newblock {\em New Astronomy}, 84:101519, 2021.

\bibitem{Nojiri:2006zh}
Shin'ichi Nojiri and Sergei~D. Odintsov.
\newblock {The New form of the equation of state for dark energy fluid and accelerating universe}.
\newblock {\em Phys. Lett. B}, 639:144--150, 2006.

\bibitem{Timoshkin:2023pfb}
A.~V. Timoshkin and A.~V. Yurov.
\newblock {Little Rip, Pseudo Rip and bounce cosmology from generalized equation of state in the universe with spatial curvature}.
\newblock {\em Int. J. Geom. Meth. Mod. Phys.}, 20(12):2350204, 2023.

\bibitem{Bento:2002ps}
M.~C. Bento, O.~Bertolami, and A.~A. Sen.
\newblock {Generalized Chaplygin gas, accelerated expansion and dark energy matter unification}.
\newblock {\em Phys. Rev. D}, 66:043507, 2002.

\bibitem{Nojiri:2003jn}
Shin'ichi Nojiri and Sergei~D. Odintsov.
\newblock {DeSitter brane universe induced by phantom and quantum effects}.
\newblock {\em Phys. Lett. B}, 565:1--9, 2003.

\bibitem{Nojiri:2003vn}
Shin'ichi Nojiri and Sergei~D. Odintsov.
\newblock {Quantum de Sitter cosmology and phantom matter}.
\newblock {\em Phys. Lett. B}, 562:147--152, 2003.

\bibitem{Nojiri:2004ip}
Shin'ichi Nojiri and Sergei~D. Odintsov.
\newblock {Quantum escape of sudden future singularity}.
\newblock {\em Phys. Lett. B}, 595:1--8, 2004.

\bibitem{Nojiri:2004pf}
Shin'ichi Nojiri and Sergei~D. Odintsov.
\newblock {The Final state and thermodynamics of dark energy universe}.
\newblock {\em Phys. Rev. D}, 70:103522, 2004.

\bibitem{Nojiri:2005sr}
Shin'ichi Nojiri and Sergei~D. Odintsov.
\newblock {Inhomogeneous equation of state of the universe: Phantom era, future singularity and crossing the phantom barrier}.
\newblock {\em Phys. Rev. D}, 72:023003, 2005.

\bibitem{Nojiri:2003ft}
Shin'ichi Nojiri and Sergei~D. Odintsov.
\newblock {Modified gravity with negative and positive powers of the curvature: Unification of the inflation and of the cosmic acceleration}.
\newblock {\em Phys. Rev. D}, 68:123512, 2003.

\bibitem{Trivedi:2023zlf}
Oem Trivedi.
\newblock {Recent advances in cosmological singularities}.
\newblock 9 2023.

\bibitem{deHaro:2023lbq}
Jaume de~Haro, Shin'ichi Nojiri, S.~D. Odintsov, V.~K. Oikonomou, and Supriya Pan.
\newblock {Finite-time cosmological singularities and the possible fate of the Universe}.
\newblock {\em Phys. Rept.}, 1034:1--114, 2023.

\bibitem{pasqua2014study}
Antonio Pasqua and Surajit Chattopadhyay.
\newblock A study on the new agegraphic dark energy model in the framework of generalized uncertainty principle for different scale factors.
\newblock {\em International Journal of Theoretical Physics}, 53(2):495--509, 2014.

\bibitem{Myung:2007pn}
Yun~Soo Myung.
\newblock {Instability of holographic dark energy models}.
\newblock {\em Phys. Lett. B}, 652:223--227, 2007.

\bibitem{bhattacharjee2020interacting}
Snehasish Bhattacharjee.
\newblock Interacting tsallis and renyi holographic dark energy with hybrid expansion law.
\newblock {\em Astrophysics and Space Science}, 365(6):103, 2020.

\bibitem{zubair2022thermodynamics}
Muhammad Zubair, Quratulien Muneer, and Ertan Gudekli.
\newblock Thermodynamics and stability analysis of tsallis holographic dark energy (thde) models in f (r, t) gravity.
\newblock {\em Annals of Physics}, 445:169068, 2022.

\end{thebibliography}
\bibliographystyle{unsrt}

\end{document}